\documentclass[prx,twocolumn,preprintnumbers,superscriptaddress]{revtex4} 

\usepackage{graphicx}
\usepackage{bm}
\usepackage{amsmath,hyperref}
\usepackage{amssymb}
\usepackage{amsfonts}
\usepackage{fancyhdr}
\usepackage{slashed}
\usepackage{verbatim}
\usepackage{latexsym,epsfig,bbm}
\usepackage{titlesec}
\usepackage{subfigure}

\newenvironment{psmallmatrix}
  {\left(\begin{smallmatrix}}
  {\end{smallmatrix}\right)}

\titleformat*{\subsubsection}{\itshape \normalsize}

\newcommand{\bra}[1]{\langle{#1}|}
\newcommand{\ket}[1]{|{#1}\rangle}

\usepackage{color}
\definecolor{blue}{rgb}{0,0.2,1}

\definecolor{red}{rgb}{0.9,0,0}

\definecolor{Pr}{rgb}{0.4,0.3,0.9}

\begin{document}

\title{Quantum machine learning for quantum anomaly detection}

\author{Nana Liu}
\email{nana.liu@quantumlah.org}
\affiliation{Centre for Quantum Technologies, National University of Singapore, 3 Science Drive 2, Singapore 117543}
\affiliation{Singapore University of Technology and Design, 8 Somapah Road, Singapore 487372}

\author{Patrick Rebentrost}
\email{patrick@xanadu.ai}
\address{Xanadu, 372 Richmond St W, Toronto, M5V 2L7, Canada}

\begin{abstract}
Anomaly detection is used for identifying data that deviate from `normal' data patterns. Its usage on classical data finds diverse applications in many important areas like fraud detection, medical diagnoses, data cleaning and surveillance. With the advent of quantum technologies, anomaly detection of \textit{quantum} data, in the form of quantum states, may become an important component of quantum applications. Machine learning algorithms are playing pivotal roles in anomaly detection using classical data. Two widely-used algorithms are kernel principal component analysis and one-class support vector machine. We find corresponding quantum algorithms to detect anomalies in quantum states. We show that these two quantum algorithms can be performed using resources logarithmic in the dimensionality of quantum states. For pure quantum states, these resources can also be logarithmic in the number of quantum states used for training the machine learning algorithm. This makes these algorithms potentially applicable to big \textit{quantum} data applications.  
\end{abstract}
\maketitle 
\section{Introduction}
Quantum computing has achieved success in finding algorithms that offer speed-ups to classical algorithms for certain problems like factoring and searching an unstructured data-base. It relies upon techniques like quantum phase estimation, quantum matrix inversion and amplitude amplification. These tools have recently been employed in quantum algorithms for machine learning\cite{qmlreview2017, dunjko2016quantum, dunjko2017machine, ciliberto2017quantum}, an area which has various applications across very broad disciplines. In particular, it has proved useful in data fitting and classification problems that appear in pattern recognition \cite{rebentrost2014quantum, lloyd2013quantum, wiebe2012quantum, wiebe2014quantum, schuld2016prediction, aimeur2007quantum, aimeur2013quantum}, where quantum algorithms can offer speed-ups in both the dimensionality and number of data used to train the algorithm. 

In the presence of a large amount of input data, some coming from unreliable or unfamiliar sources, it is important to be able to detect outliers in the data. This is especially relevant when few examples of outliers may be available to develop a prior expectation. These outliers can be indicative of some unexpected phenomena emerging in a system that has never been before identified, like a faulty system or a malicious intruder. The subject of anomaly detection is learning how to make an accurate identification of an outlier when training data of `normal' data are given. Machine learning algorithms for anomaly detection in classical data have been widely applied in areas as diverse as fraud detection, medical diagnoses, data cleaning and surveillance \cite{chandola2009anomaly, pimentel2014review}. 

Quantum data, which take the form of quantum states, are prevalent in all forms of quantum computation, quantum simulation and quantum communication. Anomaly detection of quantum states is thus expected to be an important direction in the future of quantum information processing and communication, in particular, over the cloud or quantum internet. Since machine learning algorithms have proved successful for anomaly detection in classical data, a natural question arises if there exist quantum machine algorithms used for detecting anomalies in quantum systems. While classical anomaly detection techniques for quantum states can be used, they are only possible by first probing the classical descriptions of these states which require state tomography, requiring a large number of measurements \cite{hara2014anomaly, hara2016quantum}. Thus, it would also be advantageous to reduce these resource overheads by using a quantum algorithm. 

In this paper we discuss quantum machine learning methods applied to the detection of anomalies in quantum states themselves, which we call \textit{quantum anomaly detection}. Given a training set of $M$ unknown quantum states, each of dimension $d$, the task is to use a quantum computer to detect outliers on new data, occurring for example due to a faulty quantum device. Our schemes also do not require quantum random access memory \cite{giovannetti2008quantum}, since the necessary data input is fully given by quantum states generated from quantum devices. In that sense, we present an instance of an algorithm for \textit{quantum learning} \cite{monras2017inductive, sentis2012quantum, bisio2010optimal, sasaki2001quantum, shahi2017binary}. 

We present two quantum algorithms for quantum anomaly detection: kernel principal component analysis (PCA) \cite{hoffmann2007kernel, shyu2003novel} and one-class support vector machine (SVM) \cite{scholkopf2000support, choi2009least}. We show that pure state anomaly detection algorithms can be performed using resources logarithmic in $M$ and $d$.  For mixed quantum states, we show how this is possible using resources logarithmic in $d$ only. We note this can also be an exponential resource reduction compared to state tomography \cite{nielsen2010quantum}. 

After introducing the classical kernel PCA and one-class SVM algorithms in Section~\ref{sec:CAD}, we develop quantum anomaly detection algorithms based on kernel PCA and one-class SVM for pure quantum states in Section~\ref{sec:pureQAD}. We generalise these algorithms for mixed quantum states in Section~\ref{sec:mixed QAD} and discuss the implications of these results in Section~\ref{sec:discussion}. \\

\section{Anomaly detection in classical machine learning} \label{sec:CAD}
Anomaly detection involves algorithms that can recognise outlier data compared to some expected or `normal' data patterns. These unusual data signatures, or \textit{anomalies}, can result from faulty systems, malicious intrusion into a system or from naturally occuring novel phenomena that are too rare to have been captured and classed with their own training sets. These algorithms generally provide proximity measures that quantify how `far' the inspected data pattern is from the `norm'. This is also closely related to the change point detection problem, which involves finding the point at which an underlying probability distribution governing an observed phenomena has changed \cite{basseville1993detection, sentis2016quantum}.

There are three broad classes of anomaly detection algorithms: supervised, unsupervised and semi-supervised \cite{chandola2009anomaly, pimentel2014review}. Supervised anomaly detection assume labelled training sets for \textit{both} `normal' and anomalous data and the usual supervised learning algorithms can be employed \footnote{For a simplified version of a supervised learning algorithm for quantum states, see \cite{shahi2017binary}. Also see \cite{shahi2017binary} for an unsupervised learning algorithm for quantum states.}. The unsupervised learning algorithms apply when neither `normal' nor anomalous data are labelled, but assume that `normal' cases occur far more frequently than anomalous ones. However, there are scenarios where the 'normal' data can be readily identified and gathered, but the anomalous data may be too scarce to form a training set. Here semi-supervised methods are required. We will focus on this latter scenario in this paper.

Anomaly detection algorithms can be applied to different types of anomalies: point, contextual anomalies and collective anomalies \cite{chandola2009anomaly, pimentel2014review}. Point anomalies are single data instances that can be classified as either `normal' data or an anomaly. They are the simplest, most widely studied type of anomaly and will be the focus of this paper. 

Contextual anomalies are individual data instances that are anomalous only with respect to a particular context, common in time-series data. Collective anomalies on the other hand are collections of data instances that are unusual only with respect to the whole data set. These latter types of anomalies in the quantum domain will be explored in future work. 
  
For point anomaly detection alone, there are many different types of algorithms based on different techniques \cite{chandola2009anomaly, pimentel2014review}. Two of these algorithms, kernel PCA \cite{hoffmann2007kernel} and one-class SVM \cite{choi2009least}, resemble most closely to existing quantum machine learning algorithms that provide speed-ups over their classical counterparts \cite{lloyd2014quantum, rebentrost2014quantum}. We will use these classical algorithms, described below, as a foundation to develop quantum kernel PCA and quantum one-class SVM algorithms to detect anomalies in quantum data. 
\subsection{Kernel PCA}
Suppose we are given a training set of $M$ vectors $\vec x_i \in \mathbbm{R}^M$. The centroid of the training data is denoted by
\begin{equation}
\vec x_c = \frac{1}{M} \sum_{i=1}^M \vec x_i, 
\end{equation}
and we can define $\vec z_i=\vec x_i - \vec x_c$ to be the centered data. The centered sample covariance matrix $C$ is then given by 
\begin{equation}
C = \frac{1}{M-1} \sum_{i=1}^M \vec z_i \vec z_i^T.
\end{equation}
Anomaly detection via kernel PCA is then performed using a proximity measure
\begin{equation} \label{eq:f0def}
f(\vec x_0) = |\vec z_0|^2 - \vec z_0^T C \vec z_0,
\end{equation}
where $\vec z_0=\vec x_0-\vec x_c$. This measure detects a difference in the distance between point $\vec x_0$ and the centroid of the training data and the variance of the training data along the direction $\vec z_0$. This can quantify how anomalous the point $\vec x_0$ is compared to the training data. A larger $f(\vec x_0)$ thus implies a more anomalous datum than a smaller $f(\vec x_0)$. 

This method also allows us to classify anomalies in non-linear feature spaces by performing the replacement $\vec x_i \to \phi(\vec x_i)$ with a feature map $\phi(\vec x_i)$. The inner products are performed in an abstract linear feature space. The inner product can be represented by a kernel function 
$k(\vec x_i,\vec x_j)$, which can be taken to be a non-linear function, for example $k(\vec x_i,\vec x_j)=\phi(\vec x_i)^T\phi(\vec x_j)$. For some kernels like the radial basis function kernels, it has been shown that anomaly detection can be more effective using kernel PCA compared to one-class SVM \cite{hoffmann2007kernel}. While for pure quantum states we will focus on the linear kernel, this analysis might also be extended to polynomial kernels \cite{rebentrost2014quantum} and radial basis function kernels \cite{chatterjee2016generalized}. 
\subsection{One-class SVM}
One-class SVM \cite{scholkopf2001estimating, muandet2013one} has also been applied to anomaly detection in classical data. For instance, it has been used as part of change point detection algorithms and detecting novel signals in a time series \cite{ma2003time}. We focus on the least-squares formulation of the one-class SVM \cite{choi2009least} (using the linear kernel for now). This involves finding a hyperplane in feature space such that it both maximises the distance from the hyperplane to the origin of the feature space, as well as minimising the least-squares distance $\xi_i$ between the data point $\vec x_i$ and the hyperplane. 

This is equivalent to extremising a Lagrangian $\mathcal{L}$
\begin{align}
&\mathcal{L}=\mathcal{L}(\vec{w}, r, \xi_i, \alpha_i)=\frac{1}{2}||\vec{w}||^2+\frac{1}{2P_T M}\sum_{i=1}^M \xi_i^2 \nonumber \\
&-r-\sum_{i=1}^M \alpha_i(\vec{w}^{\dagger}\cdot \vec{x}_i+\xi_i-r),
\end{align}
where $\vec{w}\cdot \vec{x}_i=r$ defines a hyperplane in feature space, which `best characterises' the set of training data and $r$ is a bias term. In the presence of non-zero $\xi_i$, the hyperplane we want to find is subject to the constraint $\xi_i=r-\vec{w}^{\dagger}\cdot \vec{x}_i$. The Lagrange multipliers $\alpha_i$ can be represented in vectorised form as $\vec \alpha$ and $P_T$ denotes a `threshold acceptance probability', related to the fraction of data expected to be outliers  \cite{scholkopf2001estimating}. From the standard method, one arrives at the corresponding matrix equation 
\begin{align} \label{eq:fmatrixeq}
\tilde{F}\begin{pmatrix}
-r \\
\vec \alpha
\end{pmatrix}=
\begin{pmatrix}
0 & \vec{e}^T \\
\vec{e} & K+P_T M \mathbf{1}
\end{pmatrix}
\begin{pmatrix}
-r \\
\vec \alpha
\end{pmatrix}=\begin{pmatrix}
1 \\
\vec{0}
\end{pmatrix},
\end{align}
where $K$ is a matrix whose elements are the kernel functions $K_{ij}=k(\vec x_i, \vec x_j)$ and $\vec e=(1,...,1)^T$. The matrix inversion of $\tilde{F}$ applied to $(1, \vec 0)^T$ then reveals $(-r, \alpha_i)$. Once $\alpha_i$ and $r$ are determined, one can compute a proximity measure $f(\vec{x}_0)$ from the new data $\vec{x}_0$ to the hyperplane
\begin{align} \label{eq:f0classical}
f(\vec{x}_0)=|\vec{w}^{\dagger} \cdot \vec{x}_0-r|,
\end{align}
which can also be generalised to non-linear kernels. 

It is possible to look at restricted versions of this algorithm, where the bias $r$ is set to a constant number $c$. This reduces Eq.~\eqref{eq:fmatrixeq} to $(K+P_TM\mathbf{1})\vec \alpha=c\vec e$. This should produce the same ordering of proximity measures when only the relative distance to the hyperplane is relevant. For simplicity, in the rest of the paper we focus on the case $r=c=1$ \footnote{We note that the special case $c=0$ can be interpreted as when the training data are all clustered around the origin of the feature space. However, it is not an appropriate choice, since solving for $(K+P_T M \mathbf{1})\vec \alpha=\vec 0$ would imply the positive semidefinite matrix $K$ has negative eigenvalues $-P_T M$, which is a contradiction.}.
\section{Pure state anomaly detection} \label{sec:pureQAD}
We begin with the following set-up. Suppose we are given access to both the unitaries $\{U_i\}$ and the control unitary $U_C=\sum_{i=1}^M \ket{i}\bra{i} \otimes U_i$, where $U_i \ket{0}=\ket{\psi_i}$ are the training quantum states. We also suppose that given any two unitaries $u_1$ and $u_2$, it is possible to generate a control unitary of the form $\ket{0}\bra{0}\otimes u_1+\ket{1}\bra{1}\otimes u_2$ \footnote{For example, in a photonic setting, one can act $u_1$ along one path of an interferometer and $u_2$ acting along the other path.}. The states $\ket{\psi_i}$ are labelled `normal'. If we are also given $U_0$ where $U_0\ket{0}=\ket{\psi_0}$, our task in quantum anomaly detection is to quantify how anomalous the state $\ket{\psi_0}$ is compared to the `normal' states. One scenario where this may be relevant is in the testing of manufactured quantum circuits. 


The control unitary $U_C$ \footnote{We note that $U_C$ can be implemented for example using passive linear optical elements \cite{araujo2014quantum}.} allows one to create superpositions of training states of the form $\sum_{i=1}^M \ket{\psi_i}\ket{i}$. This is required for the generation of the necessary kernel density matrices, the computation of some normalisation constants and for the anomaly classification stage.

With these assumptions, one can create multiple copies of the training states $\{\ket{\psi_i}\}$, the test state $\ket{\psi_0}$ and all the other states necessary to perform quantum anomaly detection. Requiring multiple identical copies of states is a purely quantum characteristic stemming from the no-cloning theorem and the imperfect distinguishability of non-orthogonal quantum states. For instance, multiple copies of two different states are required to improve the success probability in distinguishing between these two states, yet these multiple copies cannot be generated from a single copy due to the no-cloning constraint. 
\subsection{Quantum kernel PCA (pure state)}




\subsubsection{Algorithm using resources logarithmic in $d$} \label{sec:logdpca}
For a quantum version of the kernel PCA algorithm for anomaly detection, first we define the analogous centroid state of the training quantum data by
\begin{align} \label{eq:psicentre}
\ket{\psi_c}=\mathcal{N}_c\sum_{i=1}^M \ket{\psi_i}=\mathcal{N}_c\sum_{i=1}^M \sum_{j=1}^d (\psi_i)_j\ket{j}.
\end{align}
Here $(\psi_i)_j$ denotes the $j^{\text{th}}$ component of a vector $\vec \psi_i$. We demonstrate in Appendix~\ref{app:psic}  that the normalisation $|\mathcal{N}_c|^2=1/\sum_{i,j=1}^M \bra{\psi_i}\psi_j\rangle$ can be found using $\mathcal{O}(\log M)$ resources. We also show a method for preparing the centroid state in Appendix~\ref{app:psic}. 

We can write the centered quantum data as 
\begin{align} \label{eq:centrepure}
\ket{z_i}=\mathcal{N}_i(\ket{\psi_i}-(1/(M \mathcal{N}_c))\ket{\psi_c})=\mathcal{N}_i\sum_{j=1}^d  (z_i)_j \ket{j},
\end{align}
where $|\mathcal{N}_i|^2=1/(1+1/(M^2 |\mathcal{N}_c|^2)-(2/M)\text{Re}[\bra{\psi_i}\psi_c\rangle]/\mathcal{N}_c)$ and $(z_i)_j$ is the $j^{\text{th}}$ component of the vector $\vec z_i=\vec \psi_i-(1/M)\sum_{k=1}^M \vec \psi_k$. If the new quantum state to be classified is $\ket{\psi_0}=\sum_{i=1}^d (\psi_0)_i \ket{i}$, we can denote its centered equivalent by $\ket{z_0}=\mathcal{N}_0(\ket{\psi_0}-(1/(M \mathcal{N}_c))\ket{\psi_c})=\mathcal{N}_0\sum_{j=1}^d  (z_0)_j \ket{j}$, where $|\mathcal{N}_0|^2=1/(1+1/(M^2 |\mathcal{N}_c|^2)-(2/M)\text{Re}[\bra{\psi_0}\psi_c\rangle]/\mathcal{N}_c)$. 

The centered sampled covariance matrix can be rewritten in terms of quantum states as $C
=(1/(M-1))\sum_{i=1}^M (1/|\mathcal{N}_i|^2)\ket{z_i}\bra{z_i}$. It is also proportional to a density matrix $\mathcal{C}=C/\text{tr}(C)$, where $\text{tr}(C)=(1/(M-1))\sum_{i=1}^d\sum_{j=1}^M (z_j)_i (z_j)^*_i$. 

By rewriting Eq.~\eqref{eq:f0def} in terms of quantum states $\ket{z_0}$, $\mathcal{C}$ and dividing the total expression by $|\mathcal{N}_0|^2$, we can define a proximity measure $f(\psi_0)$, which obeys $0 \leq f(\psi_0) \leq 1$. This quantifies how anomalous the quantum state $\ket{\psi_0}$ is from the set $\{\ket{\psi_i}\}$ and can be written as
\begin{align} \label{eq:pcameasure}
&f(\psi_0)=\bra{z_0}(\mathbf{1}-C)\ket{z_0}=1-\text{tr}(C)\text{tr}\left(\mathcal{C} \ket{z_0}\bra{z_0}\right) \nonumber \\
&=1-\frac{\text{tr}\left(\mathcal{C} \ket{z_0}\bra{z_0}\right)}{|\mathcal{N}_{\chi_c}|^2 (M-1)}
=1-\frac{1}{M-1}\sum_{i=1}^M \frac{|\bra{z_i}z_0 \rangle|^2}{|\mathcal{N}_i|^2}.
\end{align}
In the first line, we use the fact that $\ket{z_0}$ is a normalised quantum state, which is true in all cases except $\ket{z_0}=0$. The special limit $\ket{z_0}=0$ corresponds to when all the training states and $\ket{\psi_0}$ are identical. In this case $f(\psi_0)=0$, indicating no anomaly, as expected. 

This expression in Eq.~\eqref{eq:pcameasure} is composed of a sum of $\mathcal{O}(M)$ inner product terms $\bra{\psi_i}\psi_{\kappa}\rangle$ where $i=1,...,M$ and $\kappa=0,...,M$.  Since the phases of these inner products are also required, the standard swap test \cite{buhrman2001quantum} is insufficient. Each inner product can be found instead using a modified swap test. For this we prepare the state $\ket{\Psi_{i\kappa}}=(1/2)[\ket{0}(\ket{\psi_i}+\zeta \ket{\psi_{\kappa}})+\ket{1}(\ket{\psi_i}-\zeta\ket{\psi_{\kappa}})]$ by applying the control unitary $\ket{0}\bra{0}\otimes U_i+\ket{1}\bra{1}\otimes U_{\kappa}$ and then a Hadamard gate onto the state $(1/\sqrt{2})(\ket{0}+\zeta \ket{1})\ket{0}$, where $\zeta$ is a complex number. The success probability of measuring $\ket{1}$ in the ancilla of state $\ket{\Psi_{i \kappa}}$ is given by $P_{i \kappa}=(1/2)(1-\text{Re}(\zeta \bra{\psi_i} \psi_{\kappa}\rangle))$. Then $\zeta=1$ and $\zeta=i$ recover respectively the real and imaginary part of $\bra{\psi_i}\psi_{\kappa}\rangle$. The final measurement outcomes of the ancilla state satisfy a Bernoulli distribution so the number of measurements $\mathcal{N}$ required to estimate $P_{i \kappa}$ to precision $\epsilon$ is $P_{i \kappa}(1-P_{i \kappa})/\epsilon^2$ and is upper bounded by $\mathcal{O}(1/\epsilon^2)$. Therefore $\mathcal{N} \sim \mathcal{O}(\text{poly}(\log d))$ is sufficient if an error of order $\mathcal{O}(1/\log d)$ is accepted. 

Each modified swap test requires $\mathcal{O}(\text{poly}(\log d))$ number of measurements and copies of $\ket{\psi_{\kappa}}$. The required normalisation constants $|\mathcal{N}_i|^2$ can be computed using inner products between the training states and $\mathcal{N}_c$, which require resources costing $\mathcal{O}(\text{poly}(M \log d))$. Thus the proximity measure $f(\psi_0)$ can be computed using resources scaling as $\mathcal{O}(\text{poly}(M \log d))$.  
\subsubsection{Algorithm using resources logarithmic in $d$ and $M$}
Next we show an alternative protocol for computing $f(\psi_0)$, now using $\mathcal{O}(\log M)$ resources. This can be achieved by preparing a superposition of the centered data. A method to generate $\mathcal{C}$ is to create the state $\ket{\chi_c}=\mathcal{N}_{\chi_c}\sum_{i=1}^M (1/\mathcal{N}_i)\ket{i} \ket{z_i}$ where $|\mathcal{N}_{\chi_c}|^2=1/\sum_{i=1}^M (1/|\mathcal{N}_i|^2)=M+1/(2M|\mathcal{N}_c|^2)$. In the reduced space of the second register
we find 
\begin{align}
&\text{tr}_1(\ket{\chi_c}\bra{\chi_c})=|\mathcal{N}_{\chi_c}|^2\sum_{i=1}^M (1/|\mathcal{N}_i|^2)\ket{z_i}\bra{z_i} \nonumber \\
&=|\mathcal{N}_{\chi_c}|^2 (M-1) C=\mathcal{C},
\end{align}
where $\text{tr}(C)=1/(|\mathcal{N}_{\chi_c}|^2 (M-1))$ since $\text{tr}(\text{tr}_1(\ket{\chi_c}\bra{\chi_c}))=1$. The proximity measure $f(\psi_0)=1-\text{tr}(C)\text{tr}(\mathcal{C}\ket{z_0}\bra{z_0})$ can be measured using a standard swap test with $\mathcal{O}(\log d)$ copies of $\ket{z_0}$ and $\mathcal{C}$. For methods of generating the states $\ket{z_0}$, $\ket{\chi_c}$ using $\mathcal{O}(\log M)$ resources see Appendices~\ref{app:z0} and~\ref{app:Cgen} respectively. 
\subsection{Quantum one-class SVM (pure state)}
Most classical data sets consist of real-valued numbers. In such a setting it is often sufficient to use the kernel function $k(\vec x_i, \vec x_j)=\vec x_i^T \vec x_j$ or its non-linear variants. For quantum states, it is beneficial to use well known fidelity measures such as $|\bra{\psi_i} \psi_j\rangle|^2$ \cite{nielsen2010quantum}. Thus, the kernel matrix corresponding to pure quantum states is defined by
\begin{align}
K=\sum_{i,j=1}^M |\bra{\psi_i} \psi_j\rangle|^2 \ket{i}\bra{j}. 
\end{align}
This kernel matrix is positive semi-definite, because the kernel function $|\bra{\psi_i} \psi_j\rangle|^2$ can be represented as an inner product of an appropriately defined feature map on the states $\ket {\psi_i}$. Thus $\mathcal{K}=K/\text{tr}(K)$ can be considered as a density matrix, where $\text{tr}(K)=M$. See Appendix~\ref{app:kernelpositive} for a derivation which also applies to mixed states.  

Similarly to the classical algorithm for one-class SVM, one requires a matrix inversion algorithm before computing the proximity measure, which we demonstrate below. We first present an algorithm that computes the relevant proximity measure via a swap test using resources logarithmic in the dimensionality. Then we show an algorithm potentially using resources logarithmic in the dimensionality and the size of training data based on a quantum matrix inversion algorithm. 

\subsubsection{Algorithm using resources logarithmic in $d$} \label{sec:puresvmpoly}
Suppose we look for an algorithm using $\mathcal{O}(\text{poly}(M \log (d)))$ resources when computing the proximity measure. Then it is possible to use the standard swap test to measure the fidelity $|\bra{\psi_i} \psi_j\rangle|^2$, which costs $\sim \mathcal{O}(\log (d))$ in resources for each pair $i,j=1,...,M$, or $\sim\mathcal{O}(M^2 \log (d))$ resources for every $i, j$. Then to find the constants $\alpha_i$ necessary to compute the proximity measure, one can perform a classical matrix inversion, at cost $\mathcal{O}(M^3)$. Although use of only $\mathcal{O}(\log M)$ resources is not achieved in this protocol, it does not require the assumption of having access to $U_C$. It is sufficient to need only the unitaries $\{U_i\}$ and $U_0$. 
If the unitary $U_C$ is available, we potentially obtain a logarithmic dependence in the number of quantum data as shown in the next section.

\subsubsection{Algorithm using resources logarithmic in $d$ and $M$}
\paragraph{Matrix inversion algorithm} \vspace{\baselineskip}

The problem is given by the linear system:
\begin{equation}
(K+P_TM\mathbf{1}) \vec{\alpha}=r\vec{e},
\end{equation} 
where we set $r=1$. 
We can convert this into its quantum counterpart beginning with
\begin{align}
(K+P_TM\mathbf{1})\ket{\vec{\alpha}} = \ket{\vec e},
\end{align}
where $\ket{\vec e}=(1/\sqrt{M})\sum_{i=1}^M \ket{i}$, $\ket{\vec{\alpha}}=\mathcal{N}_{\alpha} \sum_{i=1}^M \alpha_i \ket{i}$ and $\mathcal{N}_{\alpha} \equiv 1/\sum_{j=1}^M |\alpha_j|^2$. We can then use the quantum matrix inversion algorithm (HHL) \cite{HHL} to obtain $\ket{\vec \alpha}$ with a runtime of $\mathcal{O}(\log M)$. This algorithm requires the efficient exponentiation of $K$, which we show in Section~\ref{sec:expK}. 

The performance of the quantum matrix inversion algorithms relies on four basic conditions discussed in \cite{HHL} and also summarized in \cite{aaronson2015quantum}. First, the right hand side must be prepared efficiently. In the present case of the uniform superposition this step is easily performed via a Hadamard operation on all relevant qubits. 
Second,  
the quantum matrix inversion algorithm uses phase estimation as a subroutine and thus requires controlled operations of the unitary generated by the matrix under consideration. Here the controlled operations are made possible in the following way. We first rescale the overall matrix by the trace $ \text{tr}(K)=M$ so we are now solving the problem 
\begin{align}
 (\mathcal{K}+P_T\mathbf{1})\ket{\vec{\alpha}} = \ket{\vec e}.
\end{align}
Note that $\Vert \mathcal{K}+P_T\mathbf{1}\Vert = O(1)$ since $P_T$ is a constant.
Since $\mathcal{K}$ is a density matrix, we can simulate $\exp(-i \mathcal{K} t)$, with time $t$, via a variant of the quantum PCA algorithm \cite{lloyd2014quantum}, which we describe in Section~\ref{sec:expK}. For simulating this gate to precision $\epsilon$, this method requires $O(t^2/\epsilon)$ copies of $\mathcal{K}$, as long as $K$ is sufficiently low-rank.
The evolution $\exp(i P_T \mathbf{1}t)$ is trivial to generate.

As a third basic criterion, the condition number of the matrix should be at most $O(\log M)$. In our case the largest eigenvalue is $O(1+P_T)$ while the smallest eigenvalue is  $P_T$, since the kernel matrix is assumed to be low-rank with most eigenvalues being $0$. Thus we require $1/P_T \lesssim \mathcal{O}(\log M)$. The fourth criterion is that measuring the desired anomaly quantifier requires at most $O(\log M)$ repetitions, which we show in Section~\ref{sec:proxmeas}. \vspace{\baselineskip}

\paragraph{Exponentiating the kernel matrix} \label{sec:expK} \vspace{\baselineskip}
The kernel matrix $K$ can be related to another kernel matrix $K_0=\sum_{i,j=1}^M \ket{i}\bra{j}\bra{\psi_i} \psi_j\rangle$ which has been used in previous discussions of the quantum support vector machine \cite{rebentrost2014quantum}. Note that 
\begin{equation}
K= K_0^T \ast K_0,
\end{equation}
where $\ast$ is the element wise (or Hadamard) product of two matrices, $(A\ast B)_{ij} = A_{ij} B_{ij}$. 
This fact allows us to simulate the matrix exponential $\exp(-i \mathcal{K} t)$ by using multiple copies of the density matrix $\mathcal{K}_0=K_0/M$ in a modified quantum PCA method. With the efficiently simulable operator
\begin{eqnarray}
S'=\sum_{ j, k=1}^N |k\rangle \langle j|\otimes |j\rangle \langle k|  \otimes  |k\rangle \langle  j |,
\end{eqnarray}
we can perform
\begin{align}
&{\rm tr}_{1,2} \{ e^{- i S' \Delta t} ( \mathcal{K}_0 \otimes \mathcal{K}_0 \otimes \sigma ) e^{ i S' \Delta t} \} \nonumber \\
&=\sigma -i [(\mathcal{K}_0^T \ast \mathcal{K}_0 ) ,\sigma] \Delta t + O(\Delta t^2).
\end{align}
See Appendix~\ref{app:hadamard} for a derivation. Concatenating these steps allows us to perform $\exp(-i \mathcal{K} t)$ to error $O(t^2/n)$ using $2n$ copies of $\mathcal{K}_0$.

One preparation method for $\mathcal{K}_0$ \cite{rebentrost2014quantum} is by starting with the state $\ket{\chi}=(1/\sqrt{M})\sum_{i=1}^M \ket{i} \ket{\psi_i}$. Then the partial trace of $\ket{\chi}\bra{\chi}$ with respect to second register gives
\begin{align}
\text{tr}_{2}(\ket{\chi}\bra{\chi})=\frac{1}{M}\sum_{i,j=1}^M \ket{i}\bra{j}\bra{\psi_i}\psi_j \rangle=K_0/M=\mathcal{K}_0.
\end{align}

The state $\ket{\chi}$ can be generated by applying $U_C$ to the uniform superposition
\begin{equation}
\frac{1}{\sqrt{M}} \sum_{i=1}^M \ket{i} \ket 0 \to \frac{1}{\sqrt{M}} \sum_{i=1}^M \ket{i} \ket{\psi_i}=\ket{\chi}.
\end{equation} 
\vspace{\baselineskip}

\begin{figure}[ht!]
\centering
\includegraphics[scale=0.5]{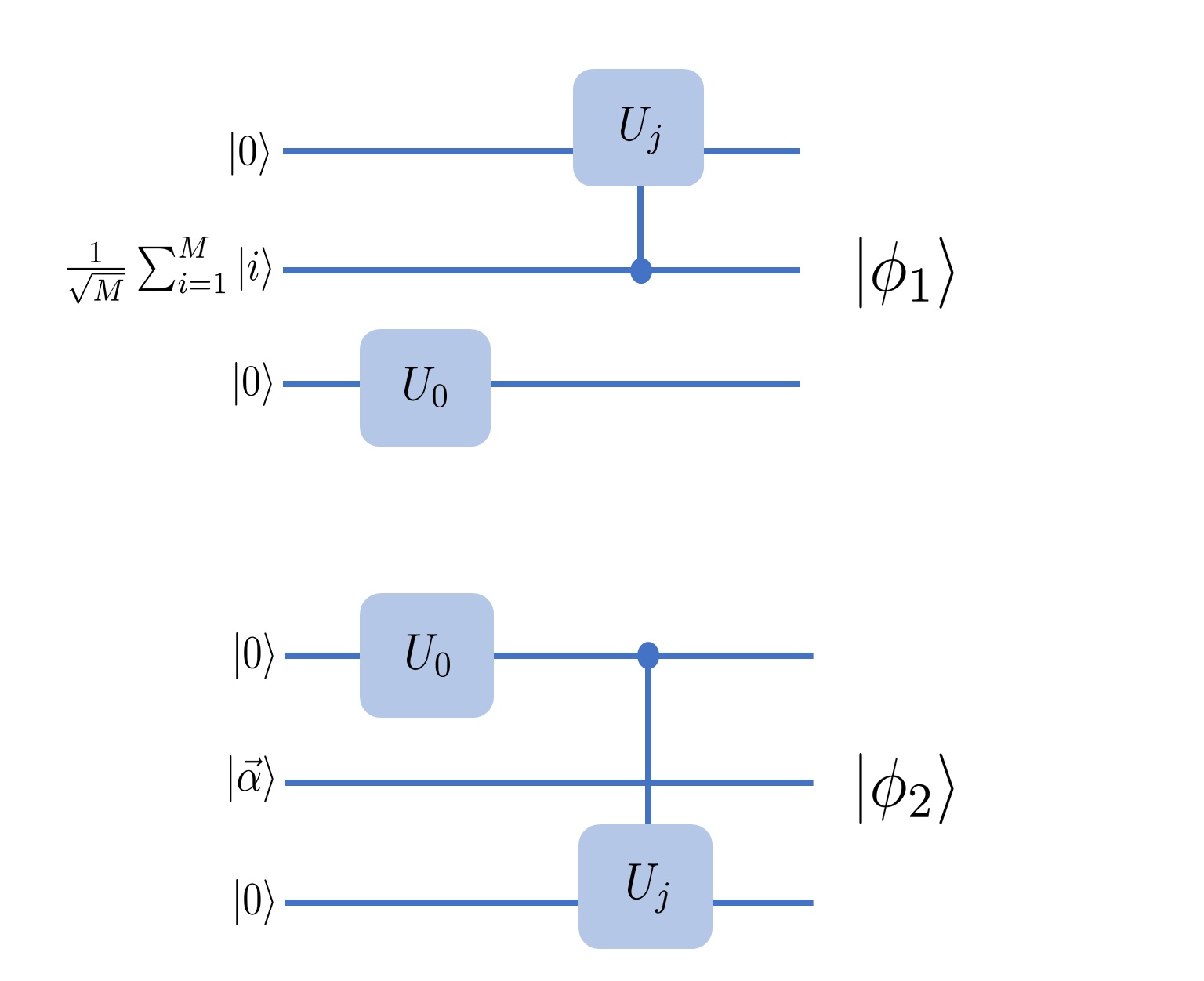}
\caption{\label{fig:phi1phi2} Quantum circuits for generating states $\ket{\phi_1}$ and $\ket{\phi_2}$. See text for more details.}
\end{figure}

\paragraph{Proximity measure computation} \label{sec:proxmeas} \vspace{\baselineskip}
The proximity measure for the pure state $\ket{\psi_0}$ when $r=1$ can be rewritten
\begin{align} \label{eq:puresvmprox}
f(\psi_0)&=|\sum_{i=1}^M \alpha_i |\bra{\psi_i}\psi_0\rangle|^2-1| \nonumber \\
&=|\bra{\phi_1}\phi_2\rangle-1|,
\end{align}
where
\begin{align}
&\ket{\phi_1}=\frac{1}{\sqrt{M}}\sum_{i=1}^M \ket{\psi_i}\ket{i}\ket{\psi_0} \nonumber \\
& \ket{\phi_2}=\frac{1}{\sqrt{M} \sum_{j=1}^M |\alpha_j|^2}\ket{\psi_0}\sum_{i=1}^M \alpha_i \ket{i} \ket{\psi_i}.
\end{align}
The state $\ket{\phi_1}$ can be generated by acting the unitary $\tilde{U}_1=\sum_{j=1}^M U_j \otimes \ket{j}\bra{j} \otimes U_0$ onto state $(1/\sqrt{M}) \sum_{i=1}^M \ket{0} \ket{i} \ket{0}$. The state $\ket{\phi_2}$ can likewise be generated by acting the unitary $\tilde{U}_2=U_0 \otimes U_C$ onto state $\ket{0}\ket{\vec{\alpha}}\ket{0}$, where $U_0 \ket{0}=\ket{\psi_0}$ and $U_i \ket{0}=\ket{\psi_i}$. See Fig.~\ref{fig:phi1phi2}. \\
 
We can compute $f(\psi_0)$ if we can then find the overlap $\bra{\phi_1}\phi_2\rangle$ in Eq.~\eqref{eq:puresvmprox} using the same kind of modified swap test in Section~\ref{sec:logdpca}. With access to the state $\ket{\tilde{\phi}_R}=(1/\sqrt{2})(\ket{0}\ket{\phi_1}+\ket{1}\ket{\phi_2})$ and applying a Hadamard on the ancilla qubit, the probability of measuring the first ancilla qubit in state $\ket{1}$ is given by $P_R=(1/2)(1-\text{Re}\bra{\phi_1}\phi_2\rangle)$. Similarly, with access with the state $\ket{\tilde{\phi}_I}=(1/\sqrt{2})(\ket{0}\ket{\phi_1}+i\ket{1}\ket{\phi_2})$ and then applying a Hadamard on the ancilla qubit, the probability of measuring the first ancilla qubit in state $\ket{1}$ is given by $P_I=(1/2)(1+\text{Im}\bra{\phi_1}\phi_2\rangle)$. The states $\ket{\tilde{\phi}_R}$ and $\ket{\tilde{\phi}_I}$ can be created by applying the control unitary $\tilde{U}_c=\ket{0}\bra{0}\otimes \tilde{U}_1 (\mathbf{1}\otimes \tilde{H}\otimes \mathbf{1})+\ket{1}\bra{1}\otimes \tilde{U}_2$ onto the states $\ket{A_R}=(1/\sqrt{2})(\ket{0}\ket{0}\ket{0}\ket{0}+\ket{1}\ket{0}\ket{\vec{\alpha}}\ket{0})$ and $\ket{A_I}=(1/\sqrt{2})(\ket{0}\ket{0}\ket{0}\ket{0}+i\ket{1}\ket{0}\ket{\vec{\alpha}}\ket{0})$ respectively, where $\tilde{H}\ket{0}=(1/\sqrt{M}) \sum_{i=1}^M \ket{i}$. See Appendix~\ref{app:controlU} for a method on generating states $\ket{A_R}$ and $\ket{A_I}$. Note that one way of implementing $\tilde{U}_c$ using $U_C$ is via a two-path interferometer with $\tilde{U}_1 (\mathbf{1}\otimes \tilde{H}\otimes \mathbf{1})$ acting along one path and $\tilde{U}_2$ acting along the other path.

An anomaly has been detected with error $\epsilon$ if we can find the value $f(\psi_0)$ to precision $\epsilon$. $f(\psi_0)$ depends only on the inner product $\bra{\phi_1}\phi_2\rangle$ and this comes from binary measurement outcomes with probabilities $P_R$ and $P_I$. Since the measurement outcomes satisfy a Bernoulli distribution, the number of measurements $\mathcal{N}_f$ to find $\text{Re}\bra{\phi_1}\phi_2\rangle$ and $\text{Im}\bra{\phi_1}\phi_2\rangle$ to precision $\epsilon$ are
\begin{align}
&\mathcal{N}_f \sim \frac{P_R(1-P_R)}{\epsilon^2}=\frac{1}{4\epsilon^2}(1-(\text{Re}\bra{\phi_1}\phi_2\rangle)^2)  \lesssim \frac{1}{4\epsilon^2} \nonumber \\
&\mathcal{N}_f \sim \frac{P_I(1-P_I)}{\epsilon^2}=\frac{1}{4\epsilon^2}(1-(\text{Im}\bra{\phi_1}\phi_2\rangle)^2)  \lesssim \frac{1}{4\epsilon^2}.
\end{align}
Thus $\mathcal{N}_f \lesssim 1/(4\epsilon^2) \sim \mathcal{O}(\log(Md))$ if the error we tolerate is of order $\mathcal{O}(1/(\log(Md))$. 
\section{Mixed state anomaly detection} \label{sec:mixed QAD}
Up to this point, we have investigated anomaly detection for pure states. One can also consider the generalised problem of anomaly detection for mixed quantum states. This can find applications in the presence of unknown noise sources or when the output of a given quantum process is designed to be a mixed state, like in NMR \cite{warren1997usefulness}. 

The performance of pure state anomaly detection is potentially logarithmic in both the dimension of the Hilbert space and the number of pure states used for training. These exponential speedups are possible as long as conditional subroutines are available to prepare the pure states and additional requirements for quantum phase estimation and matrix inversion are satisfied. However, outputs of quantum devices are often likely to be statistical mixtures of pure states. It is then natural to extend anomaly detection to mixed states as described by density matrices.

For mixed states, we have to use state similarity measures different from the pures state ones discussed above. For the kernel PCA we propose an analogous measure for mixed states and an algorithm for finding this. We show how this algorithm can be executed using $\mathcal{O}(\log d)$ resources. 

For the one-class SVM classification, one requires $K$ to be a proper kernel measure and a positive semidefinite matrix. For this we show that a quantum state fidelity measure can be used as a kernel function. We show that so long as is interested only in an algorithm that runs using $\mathcal{O}(\log d)$ resources and not necessarily efficient in $\mathcal{O}(\log M)$, the kernel matrix can be prepared. The proximity measures corresponding to mixed states can then be computed using $\mathcal{O}(\log d)$ resources.
 
However, for a big quantum data `speedup' in the number of training states, there are additional requirements to satisfy. For example, one needs an efficient way of \textit{generating} multiple copies of a density matrix proportional to the kernel matrix for the mixed states, which take a more complex form. It is unclear how these can be generated with minimal assumptions. One might also require a simple experimental procedure for computing the corresponding proximity measure. These remain interesting challenges to be explored in future work. 
\subsection{Quantum kernel PCA (mixed state)}
Suppose we want to find the proximity measure between the training states represented by density matrices $\rho_i$, $i=1,...,M$ and the test density matrix $\rho_0$. For $\kappa=0,1,...,M$, we can define these density matrices as $\rho_{\kappa}=\sum_{l=1}^N E^{{(\kappa)}}_l \ket{0}\bra{0}E^{{(\kappa)}\dagger}_l$, where $l=1,...,N$ and $E^{{(\kappa)}}_l$ are  Kraus operators obeying $\sum_{l=1}^N E^{{(\kappa)}\dagger}_lE^{{(\kappa)}}_l=\mathbf{1}$. 

We note that the pure state proximity measure in Eq.~\eqref{eq:pcameasure} is composed of $\mathcal{O}(\text{poly}(M))$ inner product terms of the form $\bra{\psi_i}\psi_{\kappa}\rangle$. We can denote a mixed state analogue of this inner product to be $\langle \rho_i, \rho_{\kappa}\rangle$. Then a proximity measure that reproduces Eq.\eqref{eq:pcameasure} in the pure state limit can be written 
\begin{align}
&f(\rho_0)=1-\frac{1}{M-1}\sum_{i=1}^M|\tilde{\mathcal{N}}_0|^2 \times \nonumber \\
&\lvert \langle \rho_i, \rho_0\rangle-\sum_{j=1}^M\frac{(\langle \rho_i, \rho_j\rangle+\langle \rho_j, \rho_0\rangle)}{M}+\frac{1}{M^2}\sum_{j,l=1}^M \langle \rho_j, \rho_l\rangle \rvert^2,
\end{align}
where $|\tilde{\mathcal{N}}_0|^2=1/(1+1/(M^2|\tilde{\mathcal{N}}_c|^2)-(2/M)\text{Re}[\sum_{i=1}^M \langle \rho_0, \rho_i\rangle/(M \tilde{\mathcal{N}}_c)]$ and $|\tilde{\mathcal{N}}_c|^2=1/\sum_{i,j=1}^M \langle \rho_i, \rho_j \rangle$, which reduce to $|\mathcal{N}_0|^2$ and $|\mathcal{N}_c|^2$ in the pure state limit. In the following, we define $\langle \rho_i, \rho_{\kappa} \rangle$ and propose a mixed state version of the modified swap test in Section~\ref{sec:logdpca} to find $\langle \rho_i, \rho_{\kappa} \rangle$.

It is important to observe that $\langle \rho_i, \rho_{\kappa} \rangle \neq \text{tr}(\rho_i \rho_{\kappa})$, since $\text{tr}(\rho_i \rho_{\kappa})$ reduces not to $\bra{\psi_i}\psi_{\kappa}\rangle$, but to the fidelity $|\bra{\psi_i}\psi_{\kappa}\rangle|^2$. Thus a simple swap test between $\rho_i$ and $\rho_{\kappa}$ is insufficient. Instead, we define
\begin{align}
\langle \rho_i, \rho_{\kappa} \rangle \equiv \sum_{l=1}^N \langle 0|E_l^{(i) \dagger} E_l^{(\kappa)}|0\rangle,
\end{align}
which obeys all the properties of an inner product. In the pure state limit, $N=1$ and the Kraus operators become unitary $E^{(\kappa)}_l \rightarrow U_{\kappa}, E^{(i)}_l \rightarrow U_{i}$, thus we can recover $\bra{\psi_i} \psi_{\kappa}\rangle$. 

To measure $\langle \rho_i, \rho_{\kappa} \rangle$, we start with an initial state $\ket{0}\bra{0}\otimes \ket{0}\bra{0}$ and apply a Hadamard to the ancilla state to get $\rho_{\text{in}}=\ket{+}\bra{+}\otimes \ket{0}\bra{0}$. Suppose we are given access to quantum operations $\mathcal{E}^{(\kappa, i)}$ such that $\mathcal{E}^{(\kappa, i)}(\rho_{\text{in}})=\sum_{l=1}^N P_l^{(\kappa, i)} \rho_{\text{in}}(P_l^{(\kappa, i)})^{\dagger}$ where $P_l^{(\kappa, i)}=\ket{0}\bra{0}\otimes E^{(\kappa)}_l+\ket{1}\bra{1}\otimes E^{(i)}_l$. Here $\sum_{l=1}^N (P_I^{(\kappa, i)})^{\dagger} P_l^{(\kappa, i)}=\mathbf{1}\otimes \mathbf{1}$ follows from completeness of the Kraus operators $E^{(\kappa)}_l$. After applying a Hadamard to the ancilla in $\mathcal{E}^{(\kappa, i)}(\rho_{\text{in}})$, the probability of finding the ancilla in state $\ket{1}$ is $(1/2)(1-\text{Re}(\sum_{l=1}^N \bra{0}E^{(i) \dagger}_l E^{(\kappa)}_l\ket{0}))$. Similarly, the imaginary components $\text{Im}(\sum_{l=1}^N \bra{0}E^{(i) \dagger}_l E^{(\kappa)}_l\ket{0})$ can be found by applying $(1/\sqrt{2})\begin{psmallmatrix} 1 & 1 \\ i & - i \end{psmallmatrix}$ to the ancilla instead of using the Hadamard gate. 

Since the final measurement outcomes of the ancilla state satisfy a Bernoulli distribution, the number of measurements $\mathcal{N}$ required to estimate $\langle \rho_i, \rho_{\kappa} \rangle$ to precision $\epsilon$ is upper bounded by $\mathcal{O}(1/\epsilon^2)$, thus $\mathcal{N} \lesssim \mathcal{O}(\text{poly}(\log d))$ if the error tolerated is of order $\mathcal{O}(1/\log d)$. Since there are $\mathcal{O}(\text{poly}(M))$ of these terms to estimate for the new proximity measure $f(\rho_0)$, the total resource cost for the algorithm is $\mathcal{O}(\text{poly}(M \log d))$.
\subsection{Quantum one-class SVM (mixed state)}
We begin by identifying the kernel matrix to the superfidelity \cite{miszczak2009sub}, which reduces to the fidelity in the pure state limit. It is possible to show that the superfidelity $F(\rho_i, \rho_j)=\text{tr}(\rho_i \rho_j)+\sqrt{1-\text{tr}(\rho_i^2)}\sqrt{1-\text{tr}(\rho_j^2)}$ between states $\rho_i$ and $\rho_j$ obey all the properties of a kernel matrix. See Appendix~\ref{app:kernelpositive}. We can then extend the proxmity measure in Eq.~\eqref{eq:puresvmprox} for pure states to mixed states 
\begin{align}
f(\rho_0)=\left|\sum_{i=1}^M \alpha_i \mathcal{F}_{i0}(\rho_{i}, \rho_0)-1\right|^2. 
\end{align}
If we can only access $\mathcal{O}(\text{poly}(M) \log (d))$ resources when computing the proximity measure, we can apply the same method as for pure states in Section~\ref{sec:puresvmpoly}. It is possible to use the same swap test, but now with mixed state inputs measuring $\text{tr}(\rho_i \rho_j)$. This again costs $\sim \mathcal{O}(\log d)$ in resources for each pair $i,j=1,...,M$, or $\sim\mathcal{O}(M^2 \log d)$ resources for every $i, j$. A classical algorithm for matrix inversion can be used, costing $\mathcal{O}(M^3)$ in resources, to find the constants $\alpha_i$. Like the pure state case, we no longer require the assumption of having access to $U_C$. 
\section{Discussion} \label{sec:discussion}
We have shown that there exist quantum machine learning algorithms for anomaly detection to detect outliers in both pure and mixed quantum states. This can be achieved using resources only logarithmic in the dimension of the quantum states. For pure states, the resources can also be logarithmic in the number of training quantum states used. 


There is a wide variety of anomaly detection algorithms based on machine learning which can be extended to the quantum realm and remain yet unexplored. These new quantum algorithms might find applications not only in identifying novel quantum phenomena, but also in secure quantum data transfer, secure quantum computation and verification over the cloud. These questions may become even more important as cloud quantum computing systems evolve into a quantum internet. 
\section*{Acknowledgements}
The authors thank J. Fitzsimons, P. Wittek, R. Munoz-Tapia, J. Calsamiglia, M. Skotiniotis, G. Sentis and A. Mantri for interesting discussions. The authors also thank P. Wittek and T. Bromley for helpful comments on the manuscript. The authors acknowledge support from the National Research Foundation and Ministry of Education, Singapore. This material is based on research supported in part by the Singapore National Research Foundation under NRF Award No. NRF-NRFF2013-01.
\appendix 
\section{Generating centroid quantum state $\ket{\psi_c}$} \label{app:psic}
Assume we are given the state $\ket{\chi}=(1/\sqrt{M})\sum_{i=1}^M \ket{\psi_i}\ket{i}$. We can apply a Hadamard operation $H$ on each of the qubits of the second register so $\ket{\chi} \rightarrow (1/M)\sum_{i=1}^M \ket{\psi_i} \sum_{j=1}^M (-1)^{ij} \ket{j}$. Then by making a projective measurement $\ket{0\dots 0}\bra{0\dots 0}$ on the second register we can recover the centroid quantum state $\ket{\psi_c}\ket{0\dots 0}$. \\

The success probability $\mathcal{P}_{\chi}$ of this measurement is
\begin{align} \label{eq:pchi}
\mathcal{P}_{\chi}&=\text{tr}(\ket{0\dots 0}\bra{0\dots 0}H^{\otimes \log M} \ket{\chi}\bra{\chi}H^{\otimes \log M} ) \nonumber \\
&=\frac{1}{M^2}\sum_{i,j=1}^M \bra{\psi_i}\psi_j\rangle. 
\end{align}
The use of $\mathcal{O}(\log M)$ Hadamard gates means that our gate resource count is $\mathcal{O}(\log M)$.

In the scenario where the covariance matrix $C$ is low rank, we expect most training samples to be rather similar, i.e., $\bra{\psi_i}\psi_j\rangle \sim \text{constant}$. This is also the case in anomaly detection where we expect the training samples to belong to a single `type' and hence expect high state fidelities between these states. In this case, $\mathcal{P}_{\chi} \sim \mathcal{O}(1)$. 

We observe that Eq.~\eqref{eq:pchi} also allows us to recover $|\mathcal{N}_c|^2=1/(M^2 \mathcal{P}_{\chi})$ using $\mathcal{O}(\log M)$ resources.  
\section{Generating centered quantum states $\ket{z_i}$ and $\ket{z_0}$} \label{app:z0}
Assume the state preparation operations as discussed in the main part. First we prepare
\begin{equation}
\frac{1}{\sqrt{2}} \left( \ket 0-\ket 1  \right)\ket {0\hdots0} \to \frac{1}{\sqrt{2}} \left( \ket 0 \ket i - \frac{1}{\sqrt M} \ket 1 \sum_{j=1}^M \ket j \right)
\end{equation}
which we obtain by performing $\ket 0 \to \ket i$ conditioned on the first ancilla being in $\ket 0$ and performing a Hadamard conditioned on the first ancilla being in $\ket 1$. Note that the later operation $\ket{0...0} \rightarrow (1/\sqrt{M})\sum_{j=1}^M \ket{j}$ requires $\mathcal{O}(\log M)$ Hadamard gates, thus the gate resource count is $\mathcal{O}(\log M)$.

Now we use $U_C$ conditioned on the non-ancilla states to prepare
\begin{align}
&\frac{1}{\sqrt{2}} \left( \ket 0 \ket i - \frac{1}{\sqrt M} \ket 1 \sum_{j=1}^M \ket j \right) \ket {0\hdots0} \nonumber \\
&\to \frac{1}{\sqrt{2}} \left( \ket 0 \ket i \ket {\psi_i} - \frac{1}{\sqrt M} \ket 1 \sum_{j=1}^M \ket j \ket {\psi_j} \right).
\end{align}
Now perform another Hadamard conditioned on the ancilla being in $\ket 1$ to arrive at
\begin{equation}
\to \frac{1}{\sqrt{2}} \left( \ket 0 \ket i \ket {\psi_i} - \frac{1}{M} \ket 1 \sum_{j,k=1}^M (-1)^{j\cdot k}\ket k \ket {\psi_j} \right).
\end{equation}
Conditioned on the ancilla being in $\ket 0$ uncompute the $\ket{i}$
\begin{equation}
\to \frac{1}{\sqrt{2}} \left( \ket 0 \ket 0 \ket {\psi_i} - \frac{1}{M} \ket 1 \sum_{j,k=1}^M (-1)^{j\cdot k}\ket k \ket {\psi_j} \right).
\end{equation}
Measure $\ket {0\cdots 0}$ in the label register to obtain
\begin{equation}
\to \frac{1}{\sqrt{1 +  \frac{1}{M^2}\sum_{j,j'=1}^M \langle \psi_{j'} \ket {\psi_j} }}  \left( \ket 0 \ket {\psi_i} - \frac{1}{M} \ket 1 \sum_{j=1}^M \ket {\psi_j} \right).
\end{equation}
The success probability of this measurement is then given by
\begin{equation}
P_{0\cdots 0}= \frac{1}{2} \left( 1 + \frac{1}{M^2}\sum_{j,j'=1}^M \langle \psi_{j'} \ket {\psi_j}  \right).
\end{equation}
As the final step measure the ancilla in $(\ket 0 + \ket 1)/\sqrt 2 $ to obtain
\begin{equation}
\to \mathcal N_i \left( \ket {\psi_i} - \frac{1}{M} \sum_{j=1}^M \ket {\psi_j} \right) \equiv \ket{z_i},
\end{equation}
where $|\mathcal{N}_i|^2=1/(1-2\text{Re}[\sum_{j=1}^M\bra{\psi_i}\psi_j\rangle]/M]+\sum_{j,k=1}^M \bra{\psi_j}\psi_k\rangle)$.
The success probability of the final measurement is given by
\begin{align}
&P_{0}^{(i)}=\frac{1}{2(1+\frac{1}{M^2}\sum_{j,j'=1}^M \langle \psi_{j'} \ket {\psi_j})} \times \nonumber \\
&\left(1-\frac{2}{M}\sum_{j=1}^M \text{Re}(\langle \psi_{i} \ket {\psi_j}) +\frac{1}{M^2}\sum_{j,j'=1}^M  \langle \psi_{j'} \ket {\psi_j}\right)
\end{align}
Thus the total probability of success is $P_{0\cdots 0} P_{0}^{(i)}$, which scales as $\sim \mathcal{O}(1)$ following a similar argument to Appendix~\ref{app:psic}. Note that in the limit where the training states are exactly the same, the state $\ket{z_i}$ does not exist, i.e., $\ket{z_i}=0$ and thus the success probability $P_0^{(i)}=0$ as expected.
\section{Generating state $\ket{\chi_c}$ for kernel PCA} \label{app:Cgen}
The state preparation of $\ket {\chi_c}$ works essentially the same way as in Appendix~\ref{app:z0}, except that we need a superposition of the label $i$. We prepare by calling the controlled state preparation operation and controlled Hadamards in the following way
\begin{align}
&\frac{1}{\sqrt{2M}} \sum_{i=1}^M \ket i \left( \ket 0 - \ket 1  \right) \ket {0\hdots0} \ket {0\hdots0} \nonumber \\
&\to\frac{1}{\sqrt{2M}}  \times \nonumber \\
&\sum_{i=1}^M \ket i \left( \ket 0 \ket {0\hdots0} \ket {\psi_i}- \frac{1}{\sqrt M} \ket 1 \sum_{j=1}^M \ket j \ket {0\hdots0} \right) \nonumber \\\\ 
&\to \frac{1}{\sqrt{2M}} \sum_{i=1}^M \ket i \left( \ket 0 \ket {0\hdots0} \ket {\psi_i}- \frac{1}{\sqrt M} \ket 1 \sum_{j=1}^M \ket j \ket {\psi_j} \right) \nonumber \\
&\to \frac{1}{\sqrt{2M}} \times \nonumber \\
& \sum_{i=1}^M \ket i \left( \ket 0 \ket {0\hdots0} \ket {\psi_i}- \frac{1}{M} \ket 1 \sum_{j,k=1}^M (-1)^{j\cdot k} \ket k \ket {\psi_j} \right).
\end{align}
Here, like in Appendices~\ref{app:psic} and~\ref{app:z0} we employed $\mathcal{O}(\log M)$ Hadamard gates for the transformation $\ket{0...0} \rightarrow (1/\sqrt{M})\sum_{j=1}^M \ket{j}$, thus our gate resource count scale as $\mathcal{O}(\log M)$.

We then measure $\ket {0\cdots 0}$ in the label register to obtain the state
\begin{equation}
\frac{\sum_{i=1}^M \ket i \left( \ket 0 \ket {\psi_i}-(1/M) \ket 1 \sum_{j=1}^M \ket {\psi_j} \right)}{\sqrt{M + 1/M \sum_{j,j'=1}^M \langle \psi_j \ket{\psi_{j'}}}}  
\end{equation}
The success probability of this measurement is given by
\begin{equation}
P_{0\cdots 0}= \frac{1}{2} +\frac{1}{2M^2} \sum_{j,j'=1}^M  \langle \psi_j \ket{\psi_{j'}}
\end{equation}
As the final step measure the ancilla in $(\ket 0 + \ket 1)/\sqrt 2 $ to obtain:
\begin{equation}
\to \mathcal N_c \sum_{i=1}^M \ket i \left( \ket {\psi_i} - \frac{1}{M} \sum_{j=1}^M \ket {\psi_j} \right) \equiv \ket{\chi_{c}},
\end{equation}
where $|\mathcal{N}_c|^2=1/(M - 1/M \sum_{j,j'=1}^M \bra{\psi_j}\psi_j'\rangle])$.
The success probability of the final measurement is then given by
\begin{equation}
P_{0}=\frac{1 - 1/M^2 \sum_{j,j'=1}^M \bra{\psi_j}\psi_j'\rangle}{1+ 1/M^2 \sum_{j,j'=1}^M \langle \psi_j \ket{\psi_{j'}}}
\end{equation}
where $P_{0} \sim \mathcal{O}(1)$ following a similar argument to Appendix~\ref{app:psic}. 
\section{Positive semi-definite kernel matrix $K$ and fidelity measure} \label{app:kernelpositive}
Let $k(A, B)$ be a kernel function, where $A$ and $B$ are matrices. This kernel function is called positive semidefinite if the kernel matrix $K_{ij}=k(\rho_i, \rho_j)$ is positive semidefinite for any training set $\{\rho_i\}$, where $\rho_i$ are density matrices.  For showing that a kernel function corresponds to a positive kernel matrix, we have by definition to show that $\sum_{ij} c_i c_j K_{ij} \geq 0$ for all vectors $\vec c$ and all training sets. Alternatively, positivity holds if there exist some feature map function $\phi(\rho_i)$ such that the kernel matrix can be defined as a matrix of inner products $K_{ij} = \phi(\rho_i)^{\dagger} \phi(\rho_j)$. 

A large number of density matrix fidelity and distance measures have been discussed \cite{jozsa1994fidelity, wang2008alternative, miszczak2009sub, puchala2011experimentally}. Here, we take the kernel function given by the \textit{superfidelity} \cite{miszczak2009sub}, defined by the symmetric function
\begin{align}
F(\rho_i, \rho_j)=\text{tr}(\rho_i \rho_j)+\sqrt{1-\text{tr}(\rho_i^2)}\sqrt{1-\text{tr}(\rho_j^2)}.
\end{align}
The matrix entries of superfidelity are all \textit{real}, since the expression $\text{tr}(\rho_i \rho_j)$ is real and positive. We can see this by observing there exist matrices $A,B$ such that $\rho_i=A^T A$ and $\rho_j=B^T B$, since $\rho_i, \rho_j$ are all positive semidefinite. Then $\text{tr}(\rho_i \rho_j)=\text{tr}((BA^T)(AB^T))=\text{tr}(C^TC)$ where $C=AB^T$ and $C^TC$ is by definition positive semidefinite. It also turns out that the super fidelity can indeed be written via a feature map, by noting that
\begin{align}
\text{tr}(\rho_i \rho_j) = \vec{\rho_i}^\dagger \vec{\rho_j},
\end{align}
where $\vec \rho$ is a vectorized form of a matrix $\rho$ by stacking the columns of $\rho$.
So the feature map is
\begin{align}
\phi(\rho_i) = \left ( \begin{array}{c} \vec{\rho}_i \\ \sqrt{1-\text{tr}(\rho_i^2)} \end{array} \right )
\end{align}
and the inner product 
\begin{align}
&\phi(\rho_i)^\dagger \phi(\rho_j) \nonumber \\
&= \vec{\rho}_i^\dagger \vec{\rho}_j +  \sqrt{1-\text{tr}(\vec \rho_i^2)}\sqrt{1-\text{tr}(\vec \rho_j^2)} = K_{ij}
\end{align}
leads to a valid positive kernel matrix. 
\section{Hadamard product and transpose} \label{app:hadamard}

Given two matrices $\rho_1$ and $\rho_2$. The task is to apply the element-wise product of these matrices as a Hamiltonian to another quantum state. This is for the purpose of simulating the kernel matrix related to the one-class SVM. 

The standard swap matrix for quantum PCA is
$S=\sum_{ j, k=1}^M |j\rangle \langle k|  \otimes  |k\rangle \langle  j |$.
Now, take the modified swap operator 
\begin{eqnarray}
S'=\sum_{ j, k=1}^M |k\rangle \langle j|\otimes |j\rangle \langle k|  \otimes  |k\rangle \langle  j | .
\end{eqnarray}
It is at most 1-sparse because each $\ket{jkj}$ gets mapped to a unique $\ket{kjk}$ while states $\ket{jkl}$ get mapped to zero for $j\neq k \neq l$. It is also efficiently computable, thus efficiently simulatable as a Hamiltonian.
With an arbitrary state $\sigma$, we can perform the following operation
\begin{equation}
{\rm tr}_{1,2} \{ e^{- i S' \Delta t} ( \rho_1 \otimes \rho_2 \otimes \sigma ) e^{ i S' \Delta t} \}.
\end{equation}
We perform the infinitesimal operation with the matrix $S'$. The trace is over the $\rho_{1}$ and $\rho_{2}$ subspaces. 
Expanding to $O(\Delta t^2)$ leads to
\begin{align}
&{\rm tr}_{1,2} \{ e^{- i S' \Delta t}( \rho_1 \otimes \rho_2 \otimes \sigma ) e^{ i S' \Delta t} \}= \nonumber \\
&1-i  {\rm tr}_{1,2} \{ S' ( \rho_1 \otimes \rho_2 \otimes \sigma ) \} \Delta t \nonumber \\
&+i {\rm tr}_{1,2} \{ ( \rho_1 \otimes \rho_2 \otimes \sigma ) S' \}\Delta t  +O(\Delta t^2).
\end{align}
The first $O(\Delta t)$ term is
\begin{align}
&{\rm tr}_{1,2}  \{ S' ( \rho_1 \otimes \rho_2 \otimes \sigma )  \} \nonumber \\
&={\rm tr}_{1,2} \{ \sum_{ j, k=1}^M |k\rangle \langle j|\otimes |j\rangle \langle k|  \otimes  |k\rangle \langle  j | ( \rho_1 \otimes \rho_2 \otimes \sigma )  \} \nonumber \\
&= \sum_{ n, m,j,k=1}^M \langle n|k\rangle \langle j|\rho_1 |n\rangle \langle m |j\rangle \langle k|  \rho_2 |m\rangle  |k\rangle \langle  j | \sigma  \nonumber \\
&= \sum_{ j,k=1}^M \langle j|\rho_1 |k \rangle  \langle k|  \rho_2 |j\rangle  |k\rangle \langle  j | \sigma  \nonumber \\
&=  (\rho_1^T \ast \rho_2 ) \sigma.
\end{align}
In the same manner we can show
\begin{align}
{\rm tr}_{1,2}  \{  ( \rho_1 \otimes \rho_2 \otimes \sigma )S'  \} 
= \sigma (\rho_1^T \ast \rho_2 ).
\end{align}
Thus in summary, we have shown that 
\begin{align}
&{\rm tr}_{1,2} \{ e^{- i S' \Delta t} ( \rho_1 \otimes \rho_2 \otimes \sigma ) e^{ i S' \Delta t} \} \nonumber \\
&= 
\sigma -i [(\rho_1^T \ast \rho_2 ) ,\sigma] \Delta t + O(\Delta t^2).
\end{align}
\section{Generating states $\ket{A_R}$ and $\ket{A_I}$} \label{app:controlU}
In Sec.~\ref{sec:proxmeas}, we required the generation of states $\ket{A_R}=(1/\sqrt{2})(\ket{0}\ket{0}\ket{0}\ket{0}+\ket{1}\ket{0}\ket{\vec\alpha}\ket{0})$ and $\ket{A_I}=(1/\sqrt{2})(\ket{0}\ket{0}\ket{0}\ket{0}+i\ket{1}\ket{0}\ket{\vec\alpha}\ket{0})$. We begin with a control-unitary operation creating the state
\begin{align} \label{eq:start}
&(\ket{0}+\ket{1})\ket{000}\ket{0} \nonumber \\
&\longrightarrow \ket{0}\ket{000}\ket{1}+\ket{1}\ket{0 \vec e\,0}\mathcal{R}\ket{0},
\end{align}
where $\ket{\vec e}=(1/\sqrt{M})\sum_{i=1}^M \ket{i}$ and $\mathcal{R}$ is the control-rotation on the ancilla (right-most qubit) used in the quantum matrix inversion algorithm \cite{HHL}. Upon measuring this ancilla in the state $\ket{1}$
\begin{align}
&\ket{0}\ket{000}\ket{1}+\ket{1}\ket{0 \vec e\,0}\mathcal{R}\ket{0} \nonumber \\
&\longrightarrow \ket{0}\ket{000}\ket{1}+\ket{1}\ket{0 \vec \alpha 0}\ket{1}.
\end{align}
Thus by eliminating the ancilla in state $\ket{1}$, we can generate $\ket{A_R}$. Similarly, if we instead begin with state $(1/\sqrt{2})(\ket{0}+i\ket{1})\ket{000}\ket{0}$ in Eq.~\eqref{eq:start}, then we can generate the state $\ket{A_I}$.
\bibliography{qmlveriref}

\end{document}